\documentclass[journal]{IEEEtran}
\usepackage{amsmath,amsfonts}
\usepackage{algorithmic}
\usepackage{algorithm}
\usepackage{array}
\usepackage[caption=false,font=footnotesize,labelfont=rm,textfont=sf]{subfig}
\usepackage{textcomp}
\usepackage{stfloats}
\usepackage{url}
\usepackage{verbatim}
\usepackage{graphicx}
\usepackage{cite}
\usepackage{xcolor}
\begin{document}

\title{Cross-Link Interference Mitigation with Over-the-Air Pilot Forwarding for Dynamic TDD}

\author{Jia-Hui~Bi, Shaoshi~Yang,~\IEEEmembership{Senior Member,~IEEE}, Xiao-Yang~Wang, Yu-Song Luo, and~Sheng~Chen,~\IEEEmembership{Life~Fellow,~IEEE}
\vspace{-5mm}
\thanks{Copyright (c) 2025 IEEE. Personal use of this material is permitted. However, permission to use this material for any other purposes must be obtained from the IEEE by sending a request to pubs-permissions@ieee.org.}
\thanks{This work was supported by Beijing Municipal Natural Science Foundation under Grant L242013. \textit{(Corresponding author: Shaoshi Yang.)}}
\thanks{J.-H. Bi, S. Yang, X.-Y. Wang and Y.-S. Luo are with the School of Information and Communication Engineering, Beijing University of Posts and Telecommunications, and the Key Laboratory of Universal Wireless Communications, Ministry of Education, Beijing 100876, China (e-mail: \{bijiahui, shaoshi.yang, wangxy\_028, yusong.luo\}@bupt.edu.cn).}
\thanks{S. Chen is with the School of Electronics and Computer Science, University of Southampton, Southampton SO17 1BJ, UK (e-mail: sqc@ecs.soton.ac.uk).}\vspace*{-4mm}
}

\maketitle

\begin{abstract}
Dynamic time-division duplex (D-TDD) aided mobile communication systems bear the potential to achieve significantly higher spectral efficiency than traditional static TDD based systems. However, strong cross-link interference (CLI) may be caused by different transmission directions between adjacent cells in D-TDD systems, thus degrading the performance. Most existing CLI mitigation schemes require sharing certain information among base stations (BSs) via backhaul links. This strategy is usually expensive and suffers high latency. Alternatively, we propose a pilot information sharing scheme based on over-the-air forwarding of the downlink pilot of the interfering BS to the interfered BS via a wireless terminal, along with a dedicated CLI channel estimation method. Simulation results demonstrate that thanks to the proposed pilot information sharing scheme the classic interference rejection combining (IRC) receiver achieves a signal detection performance highly comparable to that of the IRC detector with perfect pilot information, necessitating no information sharing among BSs via backhaul links. Furthermore, the proposed CLI channel estimation scheme reduces the impact of errors introduced by pilot forwarding, thereby improving the performance of both CLI channel estimation and signal detection.
\end{abstract}


\section{Introduction}\label{S1}
\IEEEPARstart{D}{ynamic} time-division duplex (D-TDD) allows each cell to adaptively schedule its uplink (UL) and downlink (DL) traffic based on specific demands. It is a solution to dealing with the UL and DL 
traffic asymmetry, mainly arising in dense heterogeneous
network deployments. Compared with traditional static TDD, D-TDD can significantly enhance system throughput and spectrum efficiency by offering duplex flexibility \cite{Advantages}. D-TDD has been implemented to different levels in the 4G long-term evolution (LTE) and 5G new radio (NR) systems, and is expected to be further exploited in future mobile networks.

\begin{figure}[bp]
\vspace*{-4mm}
\centerline{\includegraphics[width=\linewidth]{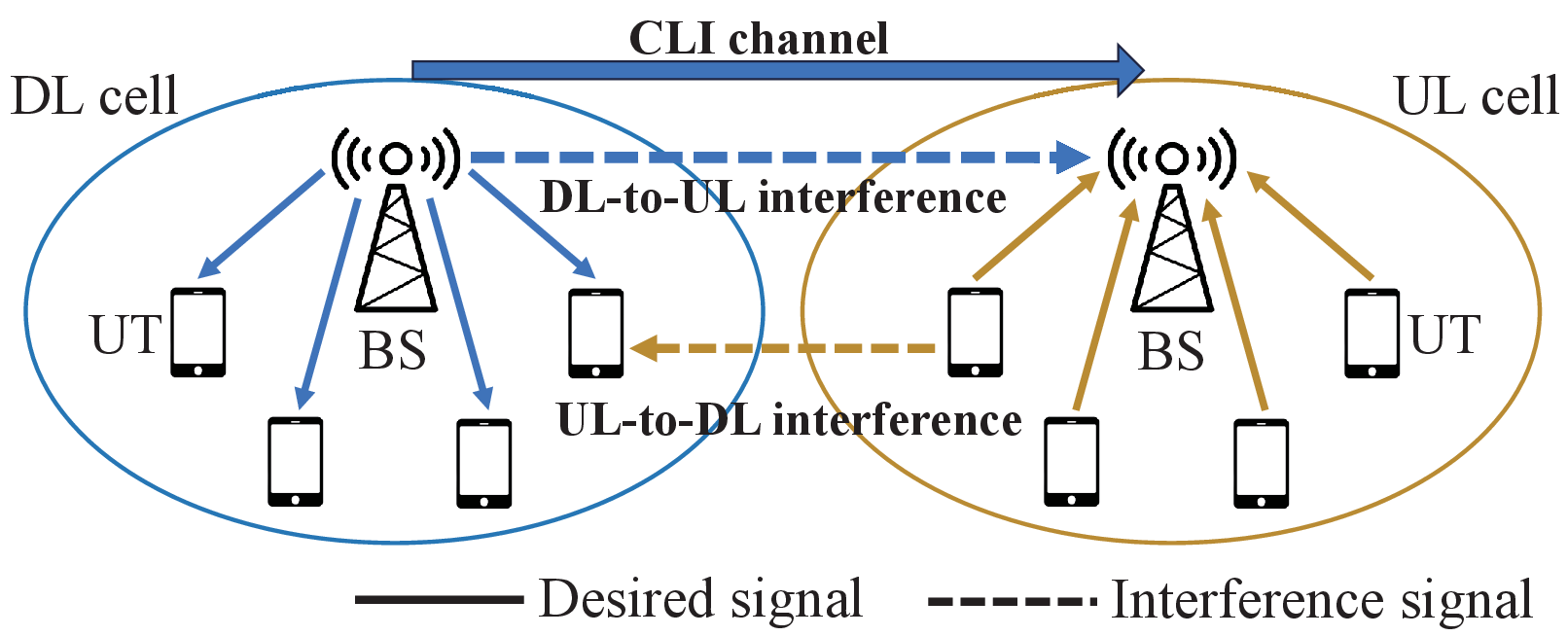}}
\vspace*{-2mm}
\caption{Illustration of CLI in a D-TDD system.}
\label{1_CLI} 
\vspace*{-0.5mm}
\end{figure}

In D-TDD systems, different cells may adopt different subframe configurations. Consequently, during certain time slots, there may be simultaneous presence of DL and UL cells within the network, causing two additional types of interference, i.e., the DL-to-UL interference and the UL-to-DL interference among adjacent cells operating in the opposite direction. These types of interference are known as cross-link interference (CLI). Specifically, the CLI caused by a DL base station (BS) to other UL BSs is referred to as DL-to-UL interference, while the CLI caused by a user terminal (UT) in a UL cell to UTs in other DL cells is referred to as UL-to-DL interference, as illustrated in Fig.~\ref{1_CLI}. CLI imposes a severe constraint on the achievable throughput performance of D-TDD systems \cite{Throughput_Analysis}, making the mitigation of CLI one of the most critical challenges in D-TDD systems.

Coordination-based schemes, including cell clustering \cite{Cell_clustering}, power control \cite{Power_control}, beamforming \cite{Beamforming} and resource allocation \cite{Resource_allocation}, mitigate CLI by decreasing the probability of its occurrence \cite{Survey}. Additionally, advanced receivers are often employed for CLI mitigation, aimed at reducing the impact of the CLI that is not avoided. In this paper, we primarily focus on advanced receiver based CLI mitigation schemes.

Interference rejection combining (IRC) detectors based on the minimum mean-square error (MMSE) criterion constitute a representative type of advanced receiver for mitigating inter-cell interference (ICI) \cite{IRC_performance} and can be employed to mitigate CLI similarly. In \cite{MMSEIRC_and_EMMSEIRC}, two linear IRC receivers for D-TDD systems were studied, i.e., the MMSE-IRC receiver and the enhanced MMSE-IRC receiver. The MMSE-IRC receiver suppresses interference based on the covariance matrix of the received signal, whereas the enhanced MMSE-IRC receiver requires the channel state information (CSI) of the CLI channel to achieve enhanced interference mitigation. Another type of advanced receiver based CLI mitigation scheme invokes interference cancellation (IC), which regenerate CLI using the estimated CSI and the transmitted data shared via backhaul links, and then cancel CLI from the received signal \cite{IC,Ding_IC_conference}. In \cite{Ding_IC_trans}, the BS-oriented partial IC scheme and the UT-oriented partial IC scheme were proposed to cancel DL-to-UL interference. Neither of them cancels all CLI signals from other DL cells but instead cancels only a portion of them to reduce the complexity and cost of the IC schemes. In recent years, machine learning has also been applied to advanced receivers for mitigating CLI. In \cite{MachineLearning}, a CLI canceller based on polynomial channel parameter estimation and two CLI cancellers based on lightweight feedforward neural network were proposed, where the nonlinearity characteristics of radio frequency chain were taken into account and excellent IC capabilities were demonstrated.

However, all the existing advanced receiver based CLI mitigation schemes require the sharing of certain information between BSs, such as demodulation reference signal (DMRS), modulation and coding scheme (MCS), radio network temporary identity, and so on, via the backhaul link \cite{Survey}, which leads to high deployment cost and communication latency. Additionally, in certain scenarios, the interfering and interfered BSs may not be able to exchange information through the backhaul link, e.g., when they belong to different mobile communication systems whose core networks are supposed to be isolated from each other.

To overcome this challenge, we propose a pilot information sharing scheme based on over-the-air (OTA) forwarding by a wireless terminal. Additionally, we introduce a dedicated CLI channel estimation scheme to reduce the impact of pilot symbol errors induced by OTA pilot forwarding on the CLI channel estimation performance. To the  best of our knowledge, this is the first advanced receiver based CLI mitigation scheme that does not require sharing information via the backhaul link. We evaluate the detection performance of the IRC receiver that employs the proposed pilot information sharing scheme. Simulation results demonstrate that the proposed OTA pilot forwarding scheme enables the IRC receiver to achieve a detection performance highly comparable to that of the IRC receiver with perfect pilot information shared via backhaul links between BSs.

\section{System Model and Problem Formulation}\label{S2}
\subsection{System Model}

In D-TDD systems, the DL-to-UL interference has a more severe impact on the system performance than the UL-to-DL interference. Therefore, this paper is dedicated to mitigating the DL-to-UL interference. We consider a multi-cell multiple-input multiple-output orthogonal frequency-division multiplexing (MIMO-OFDM) system operating in the D-TDD mode and on $W$ subcarriers. In a given time slot, the received signals in one of the UL cells are contaminated not only by white noise but also by ICI from $I$ adjacent UL cells and CLI from $J$ adjacent DL cells. We assume that the interfered BS is equipped with $R$ antennas to serve $K$ UTs. Denote the BS of the $i$th interfering UL cell by $\text{BS}_i$ and the BS of the $j$th interfering DL cell by $\text{BS}_j$. $\text{BS}_i$ is equipped with $R_i$ antennas to serve $K_i$ UTs and $\text{BS}_j$ is equipped with $N_j$ antennas to serve $M_j$ UTs. Each UT has a single antenna.

For the subcarrier $w$, let ${\mathbf{u}}_{w}\! \in\! \mathbb{C}^{K\times 1}$ be the signal vector transmitted by the $K$ UTs in the interfered cell, $\mathbf{u}_{w,i}\! \in\! \mathbb{C}^{K_i\times 1}$ be the signal vector transmitted by the $K_i$ UTs in the $i$th interfering UL cell and $\mathbf{d}_{w,j}\! \in\! \mathbb{C}^{M_j\times 1}$ be the DL data vector transmitted by $\text{BS}_j$ of the $j$th interfering DL cell to its $M_j$ UTs. The transmitted signals are normalized as $\mathbb{E} [ \mathbf{u}_{w} \mathbf{u}^{\textrm{H}}_{w} ]\! =\! \mathbf{I}_K$, $\mathbb{E}[ \mathbf{u}_{w,i} \mathbf{u}_{w,i}^{\textrm{H}} ]\! =\! \mathbf{I}_{K_i}$ and $\mathbb{E} [ \mathbf{d}_{w,j}\mathbf{d}_{w,j}^{\textrm{H}} ]\! =\! \mathbf{I}_{M_j}$, with $\mathbf{I}_{K}$ denoting the $K\! \times\! K$ identity matrix and $(\cdot )^{\textrm{H}}$ denoting the Hermitian transpose operator. The interfered BS's received signal vector ${\mathbf{y}}_{w}\! \in\! \mathbb{C}^{R\times 1}$ is given by:
\begin{equation}\label{eq1} 
	\mathbf{y}_{w} = \mathbf{H}_{\textrm{UT},w} \mathbf{u}_{w} + \sum\limits_{i} \mathbf{T}_{w,i} \mathbf{u}_{w,i} + \sum\limits_{j} \mathbf{H}_{w,j} \mathbf{W}_{w,j} \mathbf{d}_{w,j} + \mathbf{n}_{R},
\end{equation}
where $\mathbf{H}_{\textrm{UT},w}\! \in\! \mathbb{C}^{R\times K}$ is the channel frequency response (CFR) matrix at the subcarrier $w$ between the interfered BS and the UTs served by it, $\mathbf{T}_{w,i}\! \in\! \mathbb{C}^{R\times K_i}$ is the CFR matrix between the interfered BS and the UTs served by $\text{BS}_i$, $\mathbf{H}_{w,j}\! \in\! \mathbb{C}^{R\times N_j}$ is the CFR matrix between  the interfered BS and the interfering $\text{BS}_j$, and $\mathbf{W}_{w,j}\! \in\! \mathbb{C}^{N_j\times M_j}$ is the DL precoding matrix of $\text{BS}_j$, while $\mathbf{n}_{R}\! \in\! \mathbb{C}^{R\times 1}$ is a complex-valued additive white Gaussian noise (AWGN) vector with variance $\sigma_n^2$ per element, i.e., $\mathbf{n}_{R}\! \sim\! {\cal CN}\big(\mathbf{0}_{R}, \sigma_n^2 \mathbf{I}_{R}\big)$, with $\mathbf{0}_{R}$ denoting the length-$R$ zero vector.

By ignoring the weak ICI imposed by adjacent UL cells, the received signal vector at the interfered BS is expressed as:
\begin{equation}\label{model} 
	\mathbf{y}_{w} = \mathbf{H}_{\textrm{UT},w} \mathbf{u}_{w} + \mathbf{H}_{\textrm{CLI},w} \mathbf{d}_{w} + \mathbf{n}_{R},
\end{equation}
where $\mathbf{d}_{w}\! =\! \left[ \mathbf{d}_{w,1}^{\textrm{T}}, \cdots, \mathbf{d}_{w,j}^{\textrm{T}}, \cdots, \mathbf{d}_{w,J}^{\textrm{T}} \right]^{\textrm{T}}$ denotes the overall data vector transmitted by the BSs of the $J$ interfering DL cells, with $(\cdot )^{\textrm{T}}$ being the transpose operator, and $\mathbf{H}_{\textrm{CLI},w}\! =\! \left[\mathbf{H}_{w,1} \mathbf{W}_{w,1}, \cdots, \mathbf{H}_{w,j} \mathbf{W}_{w,j}, \cdots, \mathbf{H}_{w,J} \mathbf{W}_{w,J} \right]$ denotes the equivalent channel or transfer matrix between the interfering BSs and the interfered BS. In this paper, this equivalent channel is referred to as the CLI channel. Let $M=\sum\limits_j {{M_j}}$, then we have $\mathbf{H}_{\textrm{CLI},w}\! \in\! \mathbb{C}^{R\times M}$.

\subsection{IRC-Based CLI Mitigation}\label{S2.2}

The task of an advanced receiver in D-TDD systems is to recover $\mathbf{u}_{w}$ from $\mathbf{y}_{w}$.
The MMSE-IRC detector, which aims to minimize the mean-square error (MSE) between $\mathbf{u}_{w}$ and its estimate \cite{IRC}, can effectively mitigate CLI and achieve satisfactory detection performance. Let $\hat{\mathbf{H}}_{\textrm{UT},w}$ be the estimate of $\mathbf{H}_{\textrm{UT},w}$, and $\hat{\mathbf{H}}_{\textrm{CLI},w}$ be the estimate of ${\mathbf{H}}_{\textrm{CLI},w}$. The estimated signals of the MMSE-IRC detector are obtained as:
\begin{equation}\label{IRC} 
	\hat{\mathbf{u}}_{w} = \hat{\mathbf{H}}_{\textrm{UT},w}^{\textrm{H}} \mathbf{Q}_{w}^{ - 1} \mathbf{y}_{w},
\end{equation}
where $(\cdot )^{-1}$ denotes the inverse operator, and $\mathbf{Q}_{w}$ is the received signal's covariance matrix given by
\begin{equation}\label{eq5} 
	\mathbf{Q}_{w} = \hat{\mathbf{H}}_{\textrm{UT},w} \hat{\mathbf{H}}_{\textrm{UT},w}^{\textrm{H}} + \hat{\mathbf{H}}_{\textrm{CLI},w} \hat{\mathbf{H}}_{\textrm{CLI},w}^{\textrm{H}} + \sigma _n^2 \mathbf{I}_R.
\end{equation}

In this paper, we employ the MMSE-IRC algorithm to enable signal detection under severe CLI.

\subsection{Channel Estimation Methods}

Clearly, the estimation of the CLI channel is crucial for effective CLI mitigation. The estimation of the CLI channel can be directly performed using classical channel estimation schemes developed for MIMO-OFDM systems. The CLI channel can be regarded as $RM$ subchannels, each estimated separately, and the following discussion in this subsection focuses on the estimation methods for a single subchannel.

Let $x_{1}, x_{2}, \dots, x_{P}$ denote the $P$ pilot symbols used for estimating this subchannel, with their corresponding subcarrier indices given by $c_1, c_2, \dots, c_P$. The CFR of subcarrier $c$ in this subchannel is denoted by $g_c$. Then, the corresponding received signal $\mathbf{z}\! \in\! \mathbb{C}^{P\times 1}$ is given by:  
\begin{equation}\label{estimation}
    \mathbf{z} = \mathbf{X} \mathbf{g} + \mathbf{n}_{P},
\end{equation}  
where $\mathbf{X}=\textrm{diag}(x_{1}, x_{2}, \dots, x_{P})$ is a diagonal matrix, and $\mathbf{g}=[g_{c_1},g_{c_2},\cdots,g_{c_P}]^{\textrm{T}}$.

To estimate $\mathbf{g}$, we consider two classical algorithms: least-square (LS) and linear MMSE (LMMSE) \cite{Estimation}. The CFRs of subcarriers that have no pilot symbols can then be obtained through interpolation based on $\mathbf{g}$.

\subsubsection{LS Channel Estimation}
The LS channel estimation can be expressed as:  
\begin{equation}
    \hat{\mathbf{g}}_{\textrm{LS}} = \mathbf{X}^{-1} \mathbf{z}.
\end{equation}

\subsubsection{LMMSE Channel Estimation}
The LMMSE channel estimation can be expressed as:
\begin{equation}\label{LMMSE}
  \hat{\mathbf{g}}_{\textrm{LMMSE}} = \mathbf{R}_{\mathbf{gz}} \mathbf{R}_{\mathbf{z}}^{-1} \mathbf{z},
\end{equation}
where $\mathbf{R}_{\mathbf{gz}}=\mathbb{E}[ \mathbf{g} \mathbf{z}^{\textrm{H}} ]$ and $\mathbf{R}_{\mathbf{z}}=\mathbb{E}[ \mathbf{z} \mathbf{z}^{\textrm{H}} ]$. \eqref{LMMSE} can be further simplified as:
\begin{equation}\label{LMMSE2}
  \hat{\mathbf{g}}_{\textrm{LMMSE}} = \mathbf{R}_{\mathbf{g}} \left(\mathbf{R}_{\mathbf{g}}+ \mathbf{R}_{\mathbf{n}}(\mathbf{X}^{\textrm{H}}\mathbf{X})^{-1} \right)^{-1} \hat{\mathbf{g}}_{\textrm{LS}},
\end{equation}
where $\mathbf{R}_{\mathbf{g}}=\mathbb{E}[ \mathbf{g} \mathbf{g}^{\textrm{H}} ]$, $\mathbf{R}_{\mathbf{n}}=\mathbb{E}[ \mathbf{n}_P \mathbf{n}_P^{\textrm{H}} ]$. Let $\sigma_x^2$ denote the variance of the pilot symbol. Then, \eqref{LMMSE2} simplifies to \cite{Estimation}:

\begin{equation}
  \hat{\mathbf{g}}_{\textrm{LMMSE}} = \mathbf{R}_{\mathbf{g}} (\mathbf{R}_{\mathbf{g}}+ \frac{\sigma_n^2}{\sigma_x^2} \mathbf{I}_P)^{-1} \hat{\mathbf{g}}_{\textrm{LS}}.
\end{equation}

\section{Proposed Pilot Sharing Scheme and CLI Channel Estimation Scheme}\label{S3}

\subsection{Proposed DL Pilot Information Sharing Scheme}\label{S3.1}

As seen from the above exposition, the interfered BS requires the DL pilot information from the interfering BSs to perform CLI channel estimation. Existing schemes typically share DL pilot information among BSs via backhaul links, which increases deployment cost and communication latency. Furthermore, a usable backhaul link may not always be available between an interfering BS and an interfered BS. Therefore, we propose a DL pilot information sharing scheme to address the issue of CLI mitigation under such a scenario.

\begin{figure}[bp]
\vspace*{-2mm}
\centerline{\includegraphics[width=\linewidth]{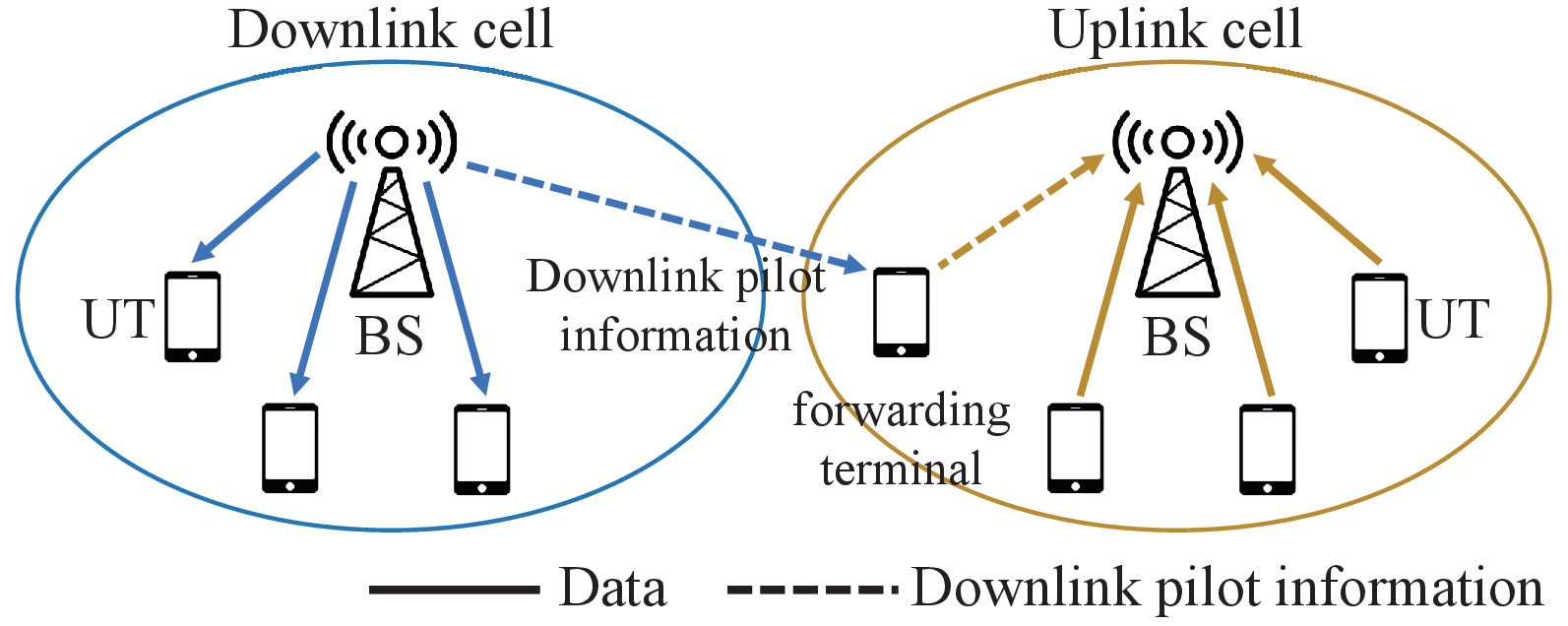}}
\vspace*{-2mm}
\caption{Illustration of the proposed DL pilot information sharing scheme.}
\label{schematic diagram} 
\vspace*{-0.5mm}
\end{figure}

\begin{figure}[tbp]
\centerline{\includegraphics[width=0.5\linewidth]{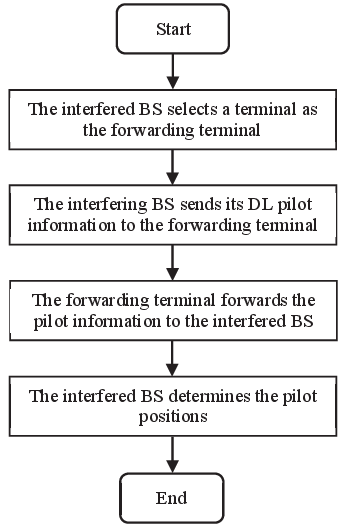}}
\vspace*{-2mm}
\caption{Flow chart of the proposed DL pilot information sharing scheme.}
\label{flow chart} 
\vspace*{-3mm}
\end{figure}

The idea of the proposed pilot information sharing scheme is that the interfering BS transmits its DL pilot information to a wireless terminal (referred to as the forwarding terminal). The forwarding terminal then forwards the pilot information to the interfered BS, as illustrated in Fig.~\ref{schematic diagram}. This forwarding terminal can be a dedicated terminal placed at a specific location in advance, or it can be a UT within either the interfered cell or the interfering cell, while having dual connections to both the interfered BS and the interfering BS, a feature supported by the related standards. The procedure of the proposed scheme is depicted in the flowchart of Fig.~\ref{flow chart}, and the detailed steps are elaborated as follows.

\subsubsection{Determine the Forwarding Terminal}
If a dedicated terminal has been pre-placed, it is selected as the forwarding terminal. Otherwise, an appropriate UT that is near the boundary of the adjacent cells and has dual connections to both cells is chosen as the forwarding terminal. 

\subsubsection{Transmit DL Pilot Information}
The interfering BS sends its DL pilot information to the forwarding terminal, using the subcarriers allocated specifically for this transmission. The information may include DL MCS, synchronization sequence, pilot sequence, and so forth. To reduce the cost of information forwarding, the interfered BS can pre-store possible values of the synchronization sequence and pilot sequence of the interfering BS. In this case, only the index information needs to be sent.

\subsubsection{Forward the Pilot Information}
The forwarding terminal sends the received information to the interfered BS during its UL time slot. Since the subcarriers used for forwarding are also specifically allocated, the interfering BS does not transmit DL data on these subcarriers, which means that the forwarded signal is free from CLI.

\subsubsection{Determine the Pilot Positions}
The interfered BS uses the synchronization sequence information received from the forwarding terminal to determine the positions of pilots in the CLI signal \cite{Synchronization}.

The proposed pilot information sharing scheme enables the interfered BS to obtain both the pilot sequence and pilot positions in the CLI signal, thereby allowing the estimation of the inter-BS transfer matrix. This scheme can be easily extended to scenarios where DL pilot information from multiple interfering BSs needs to be shared, by assigning a forwarding terminal to each interfering BS.

\subsection{Proposed CLI Channel Estimation Scheme}\label{S3.2}

The proposed pilot forwarding scheme consists of two stages: 1) the transmission stage, where the interfering BS transmits pilot information to the forwarding terminal, and 2) the forwarding stage, where the forwarding terminal relays the pilot information to the interfered BS. Both stages may introduce errors into the transmitted pilot information. In this subsection, we present a CLI channel estimation algorithm tailored to the proposed pilot forwarding scheme to reduce the impact of pilot errors on CLI channel estimation.

We first consider the transmission stage, by using a pilot symbol $x_{p}$ as an example. In this stage, the interfering BS transmits $x_{p}$ as a data symbol to the forwarding terminal. On the subcarrier carrying $x_{p}$, the interfering BS, equipped with $N_j$ antennas, simultaneously transmits data to $M'$ UTs, including the forwarding terminal, then the received signal vector $\mathbf{y}_{<p>}\! \in\! \mathbb{C}^{M'\times 1}$ is expressed as:
\begin{equation}
	\mathbf{y}_{<p>} =  \mathbf{H}_{<p>} \mathbf{W}_{<p>} \mathbf{x}_{<p>} + \mathbf{n}_{M'},
\end{equation}
where $\mathbf{H}_{<p>}\! \in\! \mathbb{C}^{ M' \times N_{j}}$ is the CFR matrix, $\mathbf{W}_{<p>}\! \in\! \mathbb{C}^{N_{j}\times M'}$ is the precoding matrix, $\mathbf{x}_{<p>}\! \in\! \mathbb{C}^{M'\times 1}$ denotes the transmitted signal vector and $x_{p}$ is an element of $\mathbf{x}_{<p>}$. We consider the zero forcing (ZF) precoding algorithm, i.e., $\mathbf{W}_{<p>}=\alpha\mathbf{H}_{<p>}^{\textrm H}(\mathbf{H}_{<p>}\mathbf{H}_{<p>}^{\textrm H})^{-1}$ with $\alpha$ being a scaling factor to satisfy the transmit power constraint \cite{precoding}. Then the recovered value of ${\mathbf{x}}_{<p>}$ can be expressed as:
\begin{equation}
	\hat{\mathbf{x}}_{<p>} = \frac{1}{\alpha} {\mathbf{y}}_{<p>}=\mathbf{x}_{<p>}+\frac{1}{\alpha} \mathbf{n}_{M'}.
\end{equation}

Let $\hat{x}_{p}$ denote the recovered value of $x_{p}$ at the forwarding terminal, and $n_{p,1}$ be the equivalent noise introduced during the transmission stage, i.e., $n_{p,1}=x_{p}-\hat{x}_{p}$. Then the variance of this equivalent noise is $\frac{\sigma_n^2}{\alpha^2}$.

Next, we consider the forwarding stage, again using the pilot symbol $x_{p}$ as an example. In this stage, the forwarding terminal transmits $\hat{x}_{p}$ to the interfered BS. On the subcarrier carrying $\hat{x}_{p}$, $K'$ UTs simultaneously transmit signals to the $R$-antenna interfered BS, and the received signal vector $\mathbf{y}_{(p)}\! \in\! \mathbb{C}^{R\times 1}$ is given by:
\begin{equation}
	\mathbf{y}_{(p)} = \mathbf{H}_{(p)}\mathbf{x}_{(p)} + \mathbf{n}_{R},
\end{equation}
where $\mathbf{H}_{(p)}\! \in\! \mathbb{C}^{R\times K'}$ is the CFR matrix, $\mathbf{x}_\textrm{(p)}\! \in\! \mathbb{C}^{K'\times 1}$ denotes the transmitted signal vector, and $\hat{x}_{p}$ is an element of $\mathbf{x}_{(p)}$. The interfered BS employs the ZF detection algorithm to recover $\mathbf{x}_{(p)}$. Let $\mathbf{F}_{p}=(\mathbf{H}_{(p)}^\textrm{H} \mathbf{H}_{(p)})^{-1} \mathbf{H}_{(p)}^\textrm{H}$, then the recovered value of ${\mathbf{x}}_{(p)}$ is given by: 
\begin{equation}
\hat{\mathbf{x}}_{(p)} =  \mathbf{F}_{p}\mathbf{y}_{(p)}=\mathbf{x}_{(p)}+\mathbf{F}_{p}\mathbf{n}_{R}.
\end{equation}

Let $\tilde{x}_{p}$ denote the recovered value of $\hat{x}_{p}$ at the interfered BS and $q$ be the index of $\hat{x}_{p}$ in $\mathbf{x}_{(p)}$, i.e., $\hat{x}_{p}$ is the $q$th element of $\mathbf{x}_{(p)}$. Let $n_{p,2}$ be the equivalent noise introduced during the forwarding stage, i.e., $n_{p,2}=\hat{x}_{p}-\tilde{x}_{p}$. This equivalent noise is given by $n_{p,2}=\sum\limits_{r = 1}^R {\mathbf{F}_{p}(q,r)\mathbf{n}_{R}(r)}$, where $\mathbf{F}_{p}(q,r)$ and $\mathbf{n}_{R}(r)$ denote the $(q,r)$th element of $\mathbf{F}_{p}$ and the $r$th element of $\mathbf{n}_{R}$, respectively. Let $\xi_p=\sum\limits_{r = 1}^R {\left|\mathbf{F}_{p}(q,r)\right|^2}$, then the variance of $n_{p,2}$ is $\xi_p\sigma_n^2$.

Let $\tilde{n}_{p}=x_{p}-\tilde{x}_{p}$ denote the total noise introduced by the two stages, then it can be expressed as $\tilde{n}_{p}=n_{p,1}+n_{p,2}$. Let $\tau_p=\frac{1}{\alpha^2}+\xi_p$. Therefore, the variance of $\tilde{n}_{p}$ is $\tau_p\sigma_n^2$.

Let $\tilde{\mathbf{X}}$ denote the recovered value of $\mathbf{X}$ at the interfered BS after forwarding, i.e., $\tilde{\mathbf{X}}=\textrm{diag}(\tilde{x}_{1}, \tilde{x}_{2}, \dots, \tilde{x}_{P})$. Consequently, with the proposed pilot information sharing scheme, \eqref{estimation} becomes:
\begin{equation}
  \mathbf{z} = \tilde{\mathbf{X}} \mathbf{g} + \tilde{\mathbf{N}} \mathbf{g} + \mathbf{n}_{P},
\end{equation}
where $\tilde{\mathbf{N}}=\textrm{diag}(\tilde{n}_{1}, \tilde{n}_{2}, \dots, \tilde{n}_{P})$. The interfered BS estimates the CLI channel based on $\tilde{\mathbf{X}}$, treating both $\tilde{\mathbf{N}} \mathbf{g}$ and $\mathbf{n}_{P}$ as noise. The covariance matrix of the equivalent noise, given by $\mathbf{n}'=\tilde{\mathbf{N}} \mathbf{g}+\mathbf{n}_{P}$, is:
\begin{equation}\label{RN'}
  \mathbf{R}_{\mathbf{n}'} = \mathbb{E}[\tilde{\mathbf{N}} \mathbf{g} \mathbf{g}^{\textrm{H}} \tilde{\mathbf{N}}^{\textrm{H}}] + \mathbb{E}[ \mathbf{n}_P \mathbf{n}_P^{\textrm{H}} ].
\end{equation}

By substituting  $\mathbf{R}_{\mathbf{g}}=\mathbb{E}[ \mathbf{g} \mathbf{g}^{\textrm{H}} ]$ and $\mathbb{E}[ \mathbf{n}_P \mathbf{n}_P^{\textrm{H}} ]= \sigma_n^2 \mathbf{I}_P$ into \eqref{RN'}, we obtain:
\begin{equation}
  \mathbf{R}_{\mathbf{n}'} = \textrm{diag}(\eta_1,\eta_2,\cdots,\eta_P),
\end{equation}
where $\eta_p=\left(\tau_p\mathbf{R}_{\mathbf{g}}(p,p)+1\right)\sigma_n^2$ with $\mathbf{R}_{\mathbf{g}}(p,p)$ denoting the $(p,p)$th element of $\mathbf{R}_{\mathbf{g}}$.

By replacing $\mathbf{R}_{\mathbf{n}}$ and $\mathbf{X}$ in \eqref{LMMSE2} with $\mathbf{R}_{\mathbf{n}'}$ and $\tilde{\mathbf{X}}$, respectively, the proposed CLI channel estimation scheme is formulated as: 
\begin{equation}\label{proposed}
  \hat{\mathbf{g}}_{\textrm{proposed}} = \mathbf{R}_{\mathbf{g}} \left(\mathbf{R}_{\mathbf{g}}+ \mathbf{R}_{\mathbf{n}'}(\tilde{\mathbf{X}}^{\textrm{H}}\tilde{\mathbf{X}})^{-1} \right)^{-1} \tilde{\mathbf{X}}^{-1} \mathbf{z},
\end{equation}

\section{Simulation Evaluation}\label{S4}

\subsection{Simulation System Settings}\label{S4.1}

To evaluate the proposed pilot information sharing scheme and CLI channel estimation scheme, we simulate a D-TDD system comprising an interfered UL cell and an interfering DL cell, i.e., $I\! =\! J\! =\! 1$. The simulation system's parameters are listed in Table~\ref{parameters}. In addition, all wireless channels are generated following the tapped delay line model specified in 3GPP TR 38.901 \cite{38901}. All UTs are randomly distributed within the cells. The interfering BS adopts the ZF DL precoding algorithm. Subcarriers for pilot transmission and forwarding are exclusively allocated to the forwarding terminal, i.e. $M'\! =\! K'\! =\! 1$.

\renewcommand{\arraystretch}{1.2}
\begin{table}[tbp]
\caption{Simulation System's Parameters}
\label{parameters} 
\vspace*{-2mm}
\centering
\begin{tabular}{|>{\centering}p{0.73\linewidth}|p{0.13\linewidth}<{\centering}|}
  \hline \bf{Parameter} & \bf{Value} \\
  \hline Carrier frequency & 3\,GHz \\
  \hline System bandwidth & 5.6\,MHz\\
  \hline Number of subcarriers & 144 \\
  \hline Subcarrier spacing & 30\,kHz \\
  \hline Pilot interval & 4 \\
  \hline Modulation scheme & 4QAM \\
  \hline Cell radius & 100\,m \\
  \hline Distance between the interfering and interfered BS & 150\,m \\
  \hline Shadow fading standard deviation & 3\,dB \\
  \hline Transmitting power of the interfering BS & 33\,dBm \\
  \hline Transmitting power of a UT & 20\,dBm \\
  \hline Transmitting power of the dedicated terminal & 20\,dBm \\
  \hline Number of antennas at the interfering BS & 8 \\
  \hline Number of antennas at the interfered BS & 8 \\
  \hline Number of UTs in the interfering cell & 4 \\
  \hline Number of UTs in the interfered cell & 4 \\
  \hline
\end{tabular}
\vspace*{-1mm}
\end{table}

\subsection{Simulation Results}\label{S4.2}
\begin{figure}[bp]
\vspace*{-2mm}
\centerline{\includegraphics[width=0.9\linewidth]{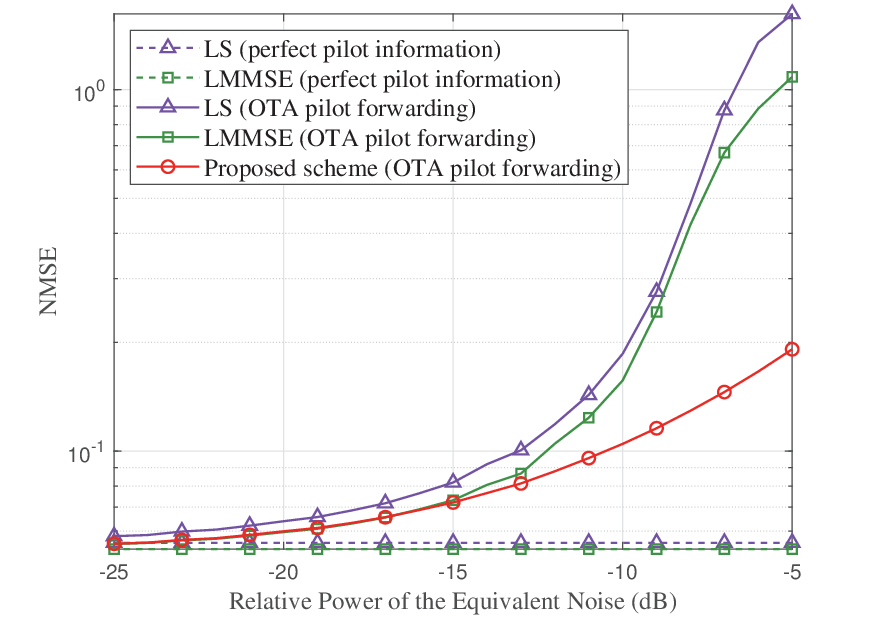}}
\vspace*{-2mm}
\caption{NMSE of different CLI channel estimation schemes as a function of the relative power of $\tilde{n}_p$.}
\label{NMSE} 
\vspace*{-0.5mm}
\end{figure}

We define the relative power of the equivalent noise $\tilde{n}_p$ introduced to the pilot symbols during pilot transmission and forwarding stages as the ratio of the average power of $\tilde{n}_p$ to that of ${x}_p$. Fig.~\ref{NMSE} presents the normalized MSEs (NMSEs) of different CLI channel estimation schemes, including LS estimation, LMMSE estimation, and the proposed estimation scheme, as a function of the relative power of $\tilde{n}_p$. LS and LMMSE estimations with perfect pilot information are used as baseline schemes. Simulation results show that when the relative power of $\tilde{n}_p$ is low, the proposed scheme exhibits no significant performance gap compared with LMMSE estimation. As the relative power increases, the NMSE values of the proposed scheme, LS estimation and LMMSE estimation all increase. However, the proposed scheme consistently achieves a lower NMSE than LS and LMMSE estimations, demonstrating its ability to reduce the impact of pilot errors on CLI channel estimation.

\begin{figure}[tbp]
\centerline{\includegraphics[width=0.9\linewidth]{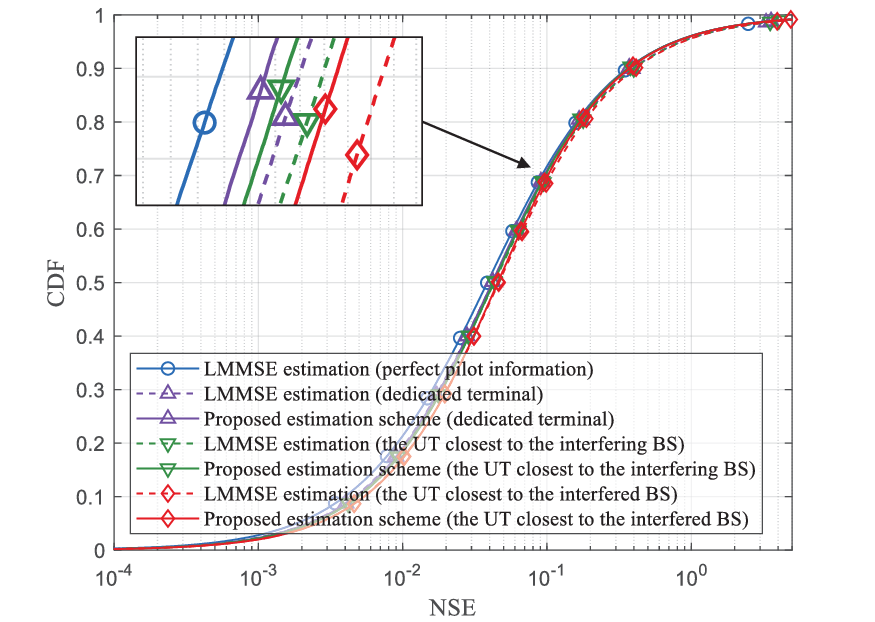}}
\vspace*{-2mm}
\caption{CDF curves of the NSEs under different CLI channel estimators and different pilot acquisition strategies.}
\label{CDF} 
\vspace*{-2mm}
\end{figure}

Both stages of the proposed pilot information sharing scheme introduce noise to the forwarded pilot symbols. We consider two strategies for selecting a UT from the interfered cell to serve as the forwarding terminal. The first strategy selects the dually connected UT closest to the interfering BS, thereby minimizing the noise introduced during the transmission stage. The second strategy selects the dually connected UT that is closest to the interfered BS, thus minimizing the noise introduced during the forwarding stage. In addition to selecting a UT as the forwarding terminal, a dedicated dually connected forwarding terminal can also be deployed specifically for forwarding pilot information. To investigate the impact of the forwarding terminal determination on the proposed pilot-sharing and channel estimation schemes, we use the normalized squared error (NSE) between the estimated and actual CFR values of the CLI channel to measure the performance of a CLI channel estimator. Fig.~\ref{CDF} presents the cumulative distribution function (CDF) curves of the NSEs corresponding to different CLI channel estimators that employ the proposed pilot information sharing scheme under various scenarios. The results corresponding to the proposed CLI channel estimation scheme and LMMSE estimation are provided for comparison. Here, the CDF of the NSEs of an LMMSE estimator with perfect pilot information is included as a benchmark. The considered scenarios include selecting the dually connected UT closest to the interfering BS, the dually connected UT closest to the interfered BS, and the dedicated dually connected terminal positioned at the midpoint of the line connecting the interfering and interfered BSs, as the forwarding terminal. Simulation results confirm that the proposed CLI channel estimation scheme consistently outperforms LMMSE estimation across all the above mentioned scenarios, aligning with the conclusion drawn from Fig.~\ref{NMSE}. Compared with using a selected UT for pilot forwarding, employing a dedicated terminal yields slightly better channel estimation performance, given both having the same transmitting power. This is because the random positions of UTs in the interfered cell lead to greater transmitting path loss than the case of the dedicated terminal. Additionally, selecting the UT closest to the interfering BS as the forwarding terminal achieves a lower NSE than selecting the UT closest to the interfered BS. In summary, deploying a dedicated forwarding terminal is recommended when feasible, despite the additional cost. If a UT from the interfered cell is employed for pilot forwarding, the UT nearest to the interfering BS should be selected.

\begin{figure}[tbp]
\centerline{\includegraphics[width=0.9\linewidth]{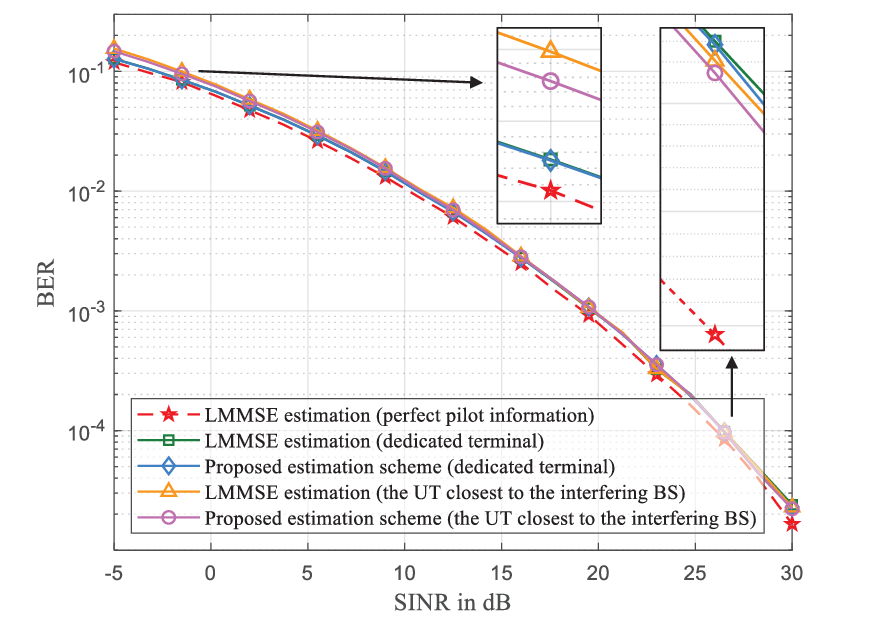}}
\vspace*{-3mm}
\caption{BERs of the MMSE-IRC receiver under different CLI channel estimation schemes and different pilot acquisition strategies, as a function of SINR.}
\label{BER1} 
\vspace*{-0.5mm}
\end{figure}

To further evaluate the performance of the proposed pilot sharing scheme and CLI channel estimation scheme, we employ the MMSE-IRC receiver introduced in Section~\ref{S2.2} to detect signals under severe CLI. Fig.~\ref{BER1} depicts the bit error rate (BER) performances of the MMSE-IRC receiver as a function of the signal-to-interference-plus-noise ratio (SINR) under different pilot acquisition strategies and CLI channel estimation schemes. Note that here SINR refers to the ratio of the power of the desired signal received by the interfered BS to the total power of AWGN and the received CLI signal. In this experiment, SINR is adjusted by configuring the transmitting power of UTs. In the considered scenario, whether a UT from the interfered cell or a dedicated terminal is employed for forwarding, the proposed pilot information sharing scheme enables the IRC receiver to achieve a BER close to that under the perfect pilot information assumption. The proposed CLI channel estimation scheme further lowers the BER.
 
\begin{figure}[tbp]
\centerline{\includegraphics[width=0.9\linewidth]{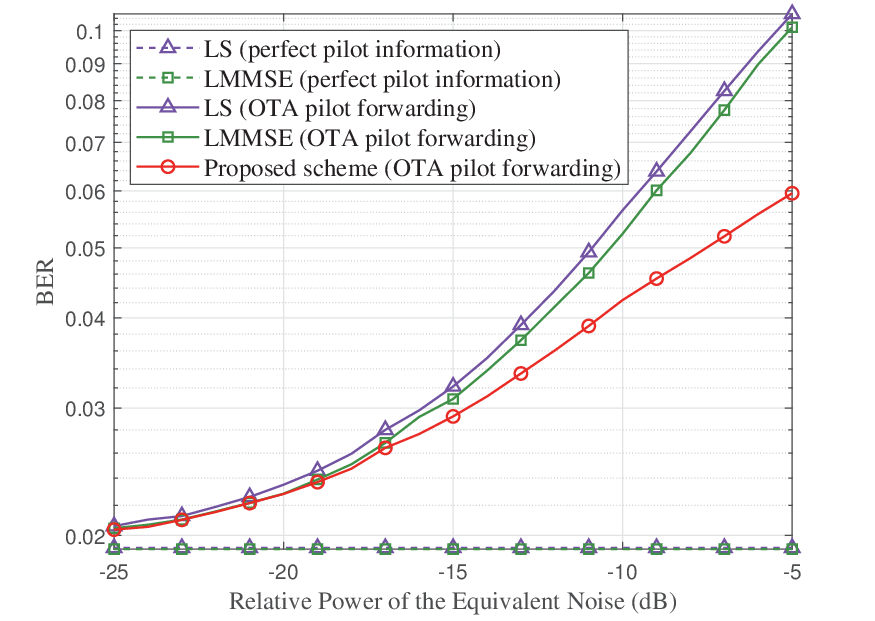}}
\vspace*{-2mm}
\caption{BERs of the MMSE-IRC receiver  under different CLI channel estimation schemes and different pilot acquisition strategies, as a function of the relative power of $\tilde{n}_p$.}
\label{BER2} 
\vspace*{-2mm}
\end{figure}

Fig.~\ref{BER2} depicts the BERs of the MMSE-IRC receiver under different CLI channel estimation schemes and different pilot acquisition strategies, as a function of the relative power of $\tilde{n}_p$. It demonstrates that when the relative noise power is small, the proposed CLI channel estimation scheme yields only a slight improvement over LMMSE. However, as the relative noise power increases, the proposed scheme gradually outperforms both LS and LMMSE, which aligns with the observed NMSE trends of Fig.~\ref{NMSE}.

\section{Conclusions}\label{S5}

We have proposed a pilot information sharing scheme based on OTA pilot forwarding, which enables the interfered BS to acquire the necessary DL pilot information of the neighboring interfering BS. As a beneficial result, advanced receivers can be implemented in the scenario where information sharing between the interfering BSs and the interfered BS via backhaul links is infeasible. Additionally, we have proposed a CLI channel estimation scheme tailored to the proposed pilot sharing framework to reduce the impact of pilot symbol errors induced by OTA pilot forwarding on CLI channel estimation accuracy. Simulation results show that although the detection performance of the MMSE-IRC receiver employing the proposed OTA pilot forwarding scheme is slightly lower than that of the MMSE-IRC receiver with perfect pilot information, they remain highly comparable. Additionally, the proposed CLI channel estimation scheme further enhances the detection performance, particularly in the scenario where the relative power of the equivalent noise introduced by OTA pilot forwarding to pilot symbols is high.

\bibliographystyle{IEEEtran}

\end{document}